# Soliton crystals in Kerr resonators

Daniel C. Cole[1,2], Erin S. Lamb[1], Pascal Del'Haye[1,†], Scott A. Diddams[1], and Scott B. Papp[1]

[1]National Institute of Standards and Technology (NIST), Boulder, CO 80305, USA
[2]Department of Physics, University of Colorado, Boulder, CO 80309, USA
[†]Present address: National Physical Laboratory (NPL), Teddington, TW11 0LW, United Kingdom

Corresponding author: daniel.cole@nist.gov

**Strongly interacting solitons confined to an optical resonator would offer unique capabilities for experiments in communication, computation, and sensing with light. Here we report on the discovery of soliton crystals in monolithic Kerr microresonators—spontaneously and collectively ordered ensembles of co-propagating solitons whose interactions discretize their allowed temporal separations. We unambiguously identify and characterize soliton crystals through analysis of their 'fingerprint' optical spectra, which arise from spectral interference between the solitons. We identify a rich space of soliton crystals exhibiting crystallographic defects, and time-domain measurements directly confirm our inference of their crystal structure. The crystallization we observe is explained by long-range soliton interactions mediated by resonator mode degeneracies, and we probe the qualitative difference between soliton crystals and a soliton liquid that forms in the absence of these interactions. Our work explores the rich physics of monolithic Kerr resonators in a new regime of dense soliton occupation and offers a way to greatly increase the efficiency of Kerr combs; further, the extreme degeneracy of the configuration space of soliton crystals suggests an implementation for a robust on-chip optical buffer.**

Optical solitons have recently found a new realization in frequency combs generated in passive, monolithic Kerr-nonlinear resonators[1] (microcombs). These microcombs are a major step forward in frequency-comb technology[2] because they enable generation of combs in platforms having low size, weight, and power requirements. When a continuous-wave pump laser is coupled into a whispering-gallery mode of a high-Q Kerr resonator, broad optical spectra are generated through cascaded four-wave mixing. With appropriate power and laser-resonator frequency detuning, the resulting fields modelock to form circulating dissipative Kerr-cavity solitons[3–10]. These solitons are pulsed excitations atop a non-zero continuous wave background, and have robust deterministic properties that may be tailored through resonator design[11,12] and tuned in real-time through manipulation of the pump laser. Microcombs based on solitons extend the range of accessible comb repetition rates and provide a route towards chip-integrated self-referenced comb technology.

Single solitons and ensembles of several co-propagating solitons have been reported in Kerr resonators constructed from a variety of crystalline and amorphous materials[3–7], with repetition rates ranging from 22 GHz[6] to 1 THz[7]. Formally equivalent to monolithic Kerr resonators are lower repetition-rate fiber-loop resonators, where generation and control of soliton ensembles has recently been explored experimentally[13–16]. These experiments were preceded by theoretical studies of soliton interactions and ensembles of solitons in quasi-CW-pumped fiber-ring resonators[17–19], where an analogy between soliton ensembles and the states of matter was introduced. This analogy has subsequently been extended to other systems including e.g. single-pass nonlinear fiber systems[20] and the harmonically mode-locked fiber laser[21,22], where a mechanism of soliton crystallization specific to that laser system was identified that is based on two different timescales of the laser gain medium[23].

In this work, we present our discovery of soliton crystallization in passive, monolithic Kerr disk resonators. The soliton crystals we present are self-organized ensembles of independent, particle-like excitations, which fill the angular domain of the resonator. These ensembles exhibit a rich configuration space supporting various crystallographic defects[24]: vacancies (Schottky defects), shifted pulses (Frenkel defects), and periodicity over multiple scales (superstructure). The crystallization mechanism that we have identified is distinct from the basic nonlinear dynamics of the system that lead to the archetypal primary comb (or Turing patterns) and modulation-instability comb (also known as spatiotemporal chaos)[8–10,25] solutions, which are extended patterns from which constituent pulses cannot be isolated and which are not degenerate with single solitons in the microcomb power-detuning phase plane[8].

Soliton crystals introduce a new regime of soliton physics into the field of microcombs. In passive CW-pumped ring resonator systems solitons exhibit locally attractive interactions at small separations that culminate in pair-annihilation or merger[14,26]; thus without, for example, manipulation of the pump laser to control these interactions[27], it has so far only been possible to observe sparsely populated ensembles of well-separated solitons in experiment. In contrast, soliton crystals lie within a new regime in which tightly packed, strongly interacting solitons fully occupy the resonator and exhibit novel collective temporal ordering. Crystallization of Kerr solitons manifests in our experiments with distinctive 'fingerprint' optical spectra arising from spectral interference between interacting solitons – similar spectra have been previously reported, but until now remained unexplained[28,29]. From the fine structure of the fingerprint spectra we deduce the configurations of soliton crystals, and we demonstrate confirmation of this process through direct measurement of the crystal configuration using ultrafast cross-correlation techniques.

A series of physical effects supported by the resonator itself is responsible for soliton crystals, and differentiates them from previously reported single- and multi-soliton microcombs. First, since the solitons are tightly packed, a soliton crystal has intracavity optical power similar to that of its experimental precursor, a time-varying chaotic intracavity waveform arising from modulation instability, which fills the angular domain of the resonator[3,8]. Hence, soliton crystals form stably through adiabatic pump-laser frequency scans, without the complex techniques used in other demonstrations of Kerr solitons to avoid dissipation of the solitons due to thermal changes. Second, destructive soliton collisions are eliminated through a naturally arising collective soliton interaction mediated by an extended background wave in the cavity. This happens because resonator mode-spectrum degeneracies locally enhance (or suppress) Kerr-comb formation, and interference of this excess light with the pump laser forms the extended wave that spontaneously orders tightly-packed soliton ensembles.

## Results

**Adiabatic generation of soliton crystals.** We generate soliton crystals in silica Kerr whispering-gallery-mode resonators with free spectral ranges (FSR) of ~26 GHz[30] and ~16.4 GHz[31]. Our experimental procedure is depicted in Fig 1a. A tapered optical fiber provides evanescent coupling of a ~30 mW telecommunications-band pump laser into a resonator whispering-gallery mode, and also out-couples the generated Kerr-soliton-crystal frequency comb. Emerging from the resonator is a pulse train with highly ordered inter-pulse timings, which we explore in detail throughout the remainder of this paper.

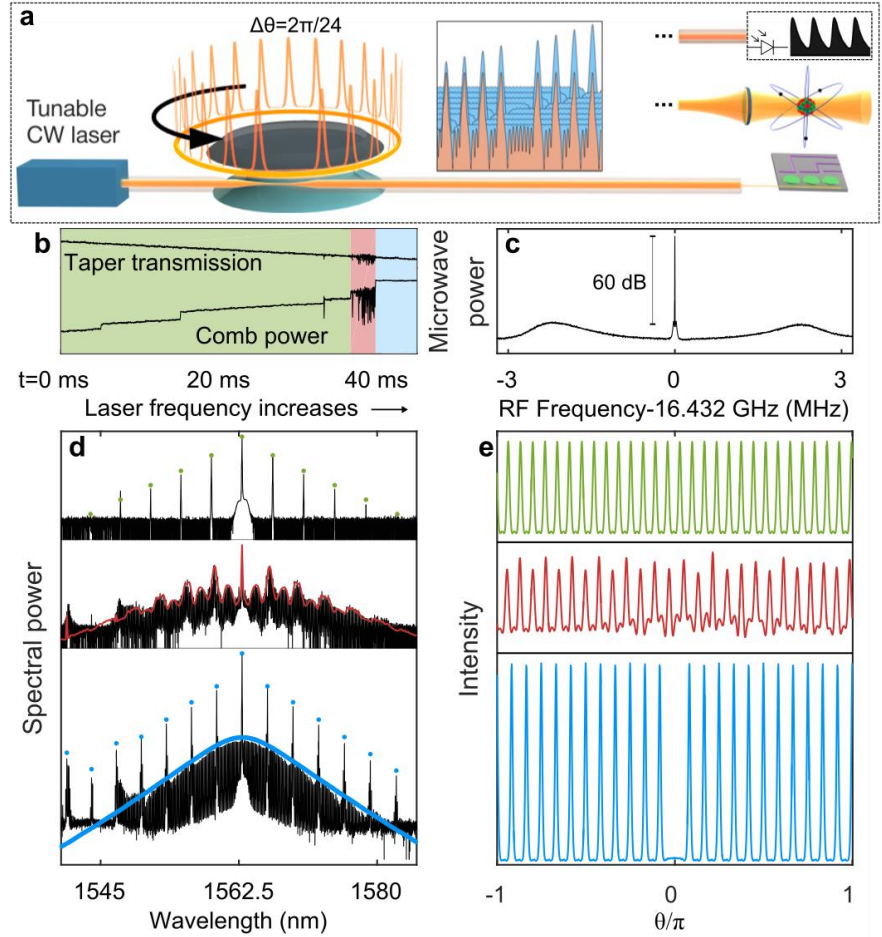

Figure 1. Generation of a soliton crystal using the Kerr nonlinearity in a $\chi_3$-nonlinear medium (here silica). (a) Depiction of generation and measurement of a crystal. Inset, plotted on logarithmic scale, depicts the summation of many single soliton waveforms to yield a crystal with a strong background oscillation. The right panel shows possible applications of the system in RF waveform generation, spectroscopy, and integrated photonics. (b) Taper transmission and comb power during crystal generation. A thermal shift of the resonance leads to the non-Lorentzian taper transmission profile. Three regimes are visible: primary comb (green), chaos (red), and crystal (blue). (c) A narrow measured RF beat indicates crystallization. (d) Progression of experimental optical spectra (black) through the three regimes, with simulations (color). Range of each plot is 100 dB. (e) Simulated time-domain waveforms corresponding to the optical spectra plotted in (d), plotted against the angular coordinate in a co-rotating frame (arbitrary vertical scale).

We form soliton crystals by using a slowly decreasing pump laser frequency scan from 'blue' to 'red' laser-cavity detuning that terminates after the soliton comb has been initiated[3]. This is similar to the experimental procedure that we reported previously in generation and characterization of phase locked microcombs[28,29]. To explore the process of Kerr-soliton-crystal formation, we record the optical power transmitted through the tapered fiber during the laser frequency scan; see Fig. 1b. By optical filtering of the pump-laser frequency, we also measure the intensity of new comb light generated by four-wave mixing. The characteristic sawtooth shape of the resonance is due to thermal bistability effects[32]. Three qualitatively different regimes can be identified in these traces: 1. Formation of initial "primary comb"[8–10,25]; 2. The chaotic waveform preceding the crystal; and 3. The Kerr soliton crystal. The continuity or near continuity of the taper transmission during the transition from chaos to crystal contrasts with the dramatic "soliton steps" that have been previously reported[3,6]. This indicates that the intracavity optical power of the crystal is similar to that of the chaos, which enables generation and stabilization of the crystals through arbitrarily slow scans of the pump laser's frequency (for example, we are able to generate soliton crystals by adjusting the pump laser's frequency by tuning the voltage across a piezo-electric crystal by hand over several seconds). We observe behavior consistent with a low-noise frequency comb following transition to the Kerr crystal not only through the fiber-taper transmission and comb-intensity signals, but also in the microwave repetition frequency signal obtained by photodetection of the entire comb; see Fig. 1c.

The most striking aspect of Kerr crystals is their optical spectra, which feature dramatic line-by-line intensity variations due to multiple-soliton interference. Figure 1d presents measurements (black traces) delineating a particular progression from primary comb to chaotic comb to crystal for the prototypical Kerr crystal in the bottom panel, which is composed of a train of 24 solitons evenly separated in time with a single vacancy. Here the optical spectrum can be understood as the destructive interference between a tightly-packed train of 24 solitons circulating the cavity, which yields the prominent comb lines spaced by 24 resonator FSR, and a single, out-of-phase soliton that both "fills in" the spectrum and produces the vacancy in the time domain. We model these spectra to obtain predictions for their intracavity intensity pattern (Fig. 1e) using the spatio-temporal Lugiato-Lefever equation[8] (LLE). The numerical results, color-coded to indicate the primary, chaotic, and crystal regimes, are highly accurate as seen through agreement over the wide logarithmic scale variations of the spectra. We note, however, that while both single solitons and few-soliton ensembles are steady-state solutions to the LLE, the 23-soliton crystal in Fig. 1d is not—when it is evolved under the LLE the solitons exhibit attractive interactions and pair-wise annihilation (see Supplement). In fact, none of the crystals we present in this Letter are stable solutions of the LLE, but they are solutions to a perturbed LLE with an altered dispersion term, as described below.

**Mechanism of Kerr soliton crystallization.** The key to understanding why soliton crystals exist can be seen in their spectra. We observe excess Kerr comb generation relative to the expected hyperbolic-secant spectrum (not shown), e.g. 6 dB near 1547 nm and 18 dB near 1541 nm in Fig. 1d, due to accidental degeneracies of the resonator mode family that supports our combs and other mode families. Such mode crossings lead to a dispersive change in the comb-resonator frequency detuning about the crossing, either increasing or decreasing the efficiency of comb formation. Mode crossing phenomena have discussed previously: in anomalous-dispersion resonators they can inhibit the formation of Kerr solitons[4,6], while they can facilitate the formation of Kerr combs in for normal dispersion[33]. In our work in the anomalous dispersion regime, mode crossings play a critical role in stabilizing attractive interactions within tightly packed, multiple-soliton ensembles. Formally, the result of a mode crossing is the incorporation of an extended background wave into the soliton waveform circulating in the resonator. This wave corresponds to the interference of excess light with the pump laser. When several of these perturbed solitons co-propagate in a resonator, they interact through their extended waves and arrange themselves such that the waves constructively interfere. Each soliton then lies at the peak of an extended background wave in the resonator, similar to predictions for bi-chromatically pumped Kerr combs[34]. Importantly, temporal separations between solitons are therefore required to be multiples of this wave's period, and the wave stabilizes the crystal against the attractive interactions discussed above. Further, the wave's amplitude, and thus the strength of the crystal against perturbations, increases with the number of co-propagating perturbed solitons. Finally, we note that a mode-crossing which predominantly affects a single frequency comb mode leads, effectively, to the injection of a single CW laser into the cavity waveform, giving this interaction infinite range at least within the assumption of single-mode perturbation.

The mechanism for soliton crystallization that we have discovered is a synthesis of and elaboration upon previously reported phenomena: it has been shown that local interactions between cavity solitons can arise through decaying oscillatory tails[35], leading to the formation of small, locally ordered soliton molecules. Further, it has been shown that the injection of an additional CW laser into a passive fiber-ring resonator can result in the generation of uniform distributions of solitons[36] – this can be viewed as CW-soliton interaction. In our experiments the 'injected' CW laser is provided naturally by the solitons themselves.

As one specific example of spontaneous crystallization driven by the extended background wave, the spectrum in the bottom of Fig. 1d exhibits excess power near optical modes 5 x 24=120 (1547 nm) and 7 x 24=168 (1541 nm), counted from the pump laser. Also visible is suppressed comb generation where the comb-resonator detuning has been increased. Here 24 FSR is the spacing of the prominent primary-comb lines arising from the modulation instability phase-matching condition, and 24 is approximately the maximum number of soliton pulses that can be seeded into the resonator. A phenomenological application of coupled-mode theory[37] could be employed to calculate the spectrum of the solitons perturbed by the mode crossing, but we find that to explain the stability of this 23 soliton crystal, and the apparently exact circumferential spacing of the pulses by $2\pi/24$ radians, it is sufficient to incorporate into the LLE a reduced comb-resonator detuning on only mode 120 or on mode 168, where the excess power is largest. Then the crystal is a steady-state solution of the resulting perturbed LLE.

**A superstructured spectrum and contrast with the soliton liquid.** With the above understanding of soliton crystallization in place, we can consider a range of crystal configurations that can be explained through this universal model. To begin, we consider a second crystal spectrum, shown in Fig. 2, which was first reported by Del'Haye et al.[29] This crystal exhibits superstructure—the soliton train is nearly periodic in a small unit cell but is modulated with a larger periodicity. This results from the frustrated uniform distribution of 16 solitons with allowed inter-soliton separations of $2\pi n/49$ radians; two solitons are spaced by $4 \times 2\pi/49$ instead of $3 \times 2\pi/49$ radians. Excess power is apparent in the spectrum at mode 49 (~1542 nm), and we simulate this crystal by phenomenologically reducing the comb-resonator detuning on mode 49 so that the experimental and simulated spectra agree. The background wave resulting from the constructive interference of the extended waves of the solitons, each having an angular period of $2\pi/49$, is visible in the plots of the simulated intensity in Figs. 2b and 2c.

To gain insight into crystal generation, we simulate laser frequency scans across the resonance that generate this crystal in the presence of the mode crossing on mode 49. Example scans are shown for the case without the mode crossing (green) and with it (blue) in Fig. 2d. In both scans, solitons emerge from chaos as the frequency of the laser is decreased. In the presence of the mode crossing, they are generated with inter-soliton separations of $2\pi n/49$ radians. A greater number of solitons emerge from chaos in the presence of the mode crossing, and this higher number helps to stabilize the crystal against thermal changes in the experiment. Further, upon continuation of the simulation, some of the solitons in the scan without the mode crossing interact attractively and pair-annihilate, while the crystallized ensemble resulting from the scan with the mode crossing remains stable indefinitely.

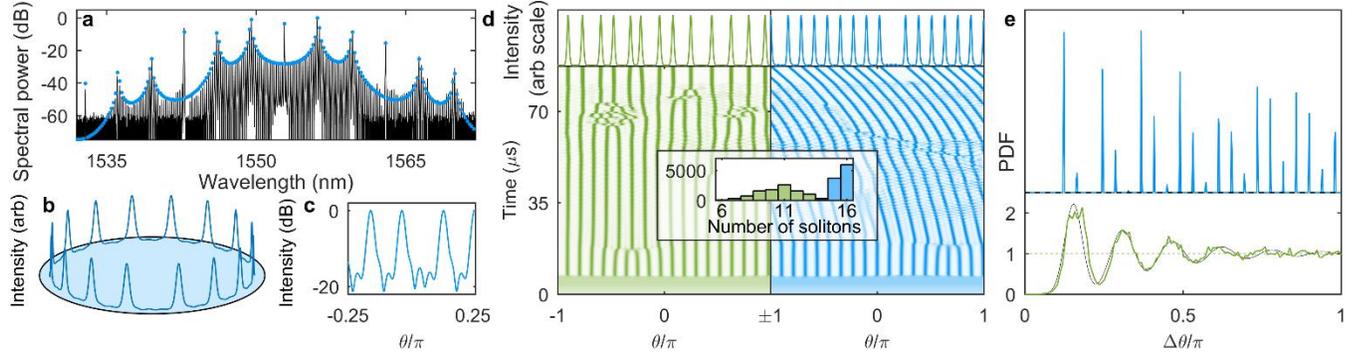

Figure 2. Investigation of a superstructured crystal. (a) Experimental (black) and simulated (blue) optical spectra. (b, c) Simulated time-domain power of 16 nearly-uniformly distributed solitons, corresponding to the simulated spectrum in (a), on linear (b) and logarithmic (c) scales. (d) Bottom: Plots of the intracavity power during a simulated scan across the resonance, without (green) and with (blue) a mode crossing. Top: Final waveforms. Inset: Histogram of the number of solitons generated in 10000 simulated scans. (e) Average pair-distribution functions for 10000 simulated scans across the resonance with (blue) and without (green) a mode crossing. The width of the peaks in the discrete pair-distribution functions is a single $\Delta\theta$ bin. The expected PDF of a simple 1-d liquid (see text) is plotted for comparison (black).

We investigate the pair-distribution function (PDF) for the soliton ensembles generated by these scans. The PDF is the probability that a soliton exists at position $\theta_o + \Delta\theta$ given that a different soliton exists at position $\theta_o$, normalized to the density, and this metric is commonly used to classify particle interactions in condensed matter physics (e.g. Ref. 38). We note that for numerically calculated discrete PDFs the absolute scaling of the PDF is not important, as it depends on the density of points. In Fig. 2e, we plot the average PDFs for 10000 simulated scans with and without the mode crossing. The result for the case with a mode crossing is sharply peaked, indicating that the allowed inter-soliton separations take on discrete values. The result for the case without the mode crossing is continuous, with a peak near the most likely nearest-neighbor separation and periodic revivals at its multiples, falling to the value of the PDF for uncorrelated soliton positions (the density) at large separations. This is exactly the expected form of the PDF for a liquid[38]. For comparison we plot a PDF (black, Fig. 2e) generated by simulation of a simple particle ensemble with mean inter-particle separation of $\Delta\theta = 0.155\pi$ and normally distributed noise on this value with standard deviation of $\sigma_{\delta\theta} = 0.18\,\Delta\theta$. Thus, with a particle labeled by $n = 0$ fixed at $\theta = 0$, the position of particle n is $\theta_n = n\Delta\theta + \sum_i^n \delta\theta_i$, with $\delta\theta_i$ the instantiations of the random variable representing the noise on the pulse spacings.

**Soliton crystal configuration space.** We observe a rich variety of soliton crystals that correspond to multiple-soliton configurations ordered according to an extended background wave; many of the optical spectra are plotted in Fig. 3. Operationally, we adjust the pump laser power to provide repeatable conditions for creating particular crystals, with more complex crystals occurring with increased laser power, which intensifies the chaos that precedes crystal generation and provides less well-ordered initial conditions. These crystals exhibit vacancies (Schottky defects)[24], Frenkel defects[24], disorder, or superstructure, or some combination thereof. A Frenkel defect consists of the shifting of a soliton in an otherwise uniform crystal. Disordered crystals are crystals in which the solitons fall on the peaks of the extended background wave, but their distribution across these peaks varies without any apparent regular order or favored period.

We highlight the crystal plotted in Fig. 3n. This crystal exhibits both superstructure, with a superlattice period of $2\pi/3$ radians, and a Frenkel defect. Three identical supercells per resonator round-trip yield a spectrum which has light in optical modes spaced by three resonator FSR, because the waveform's period has been reduced threefold.

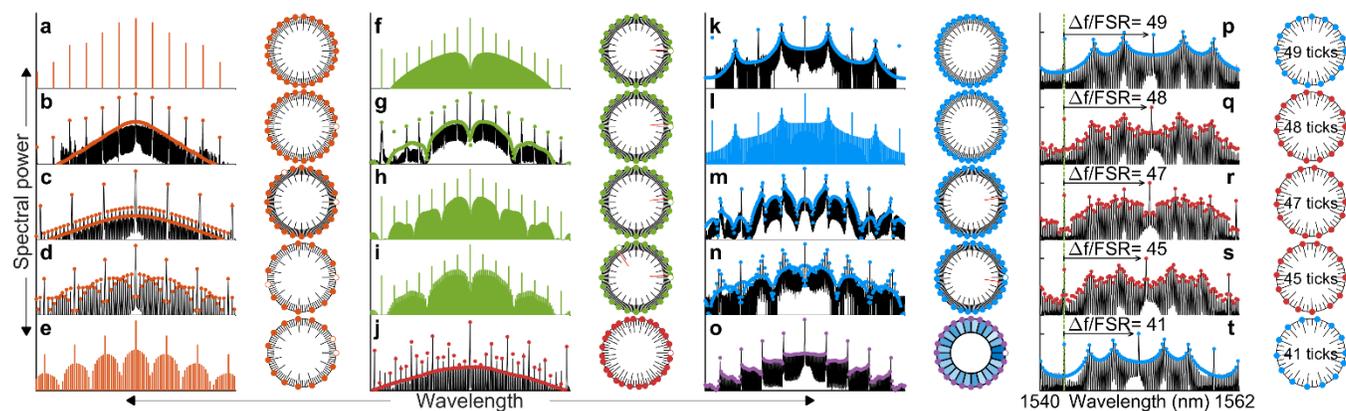

Figure 3. A taxonomy of soliton crystals. Measured optical spectra are shown in black, simulations are shown in color. Schematic depictions of the soliton distribution in the resonator are shown to the right of each spectrum. Major ticks in the schematic diagram indicate the location or expected location of a soliton. Minor ticks indicate peaks of the extended background wave due to the mode crossing. (a) A perfect soliton crystal, consisting of 25 uniformly-distributed solitons. (b-e) Soliton crystals exhibiting vacancies. (f-i) Soliton crystals exhibiting Frenkel defects. Shifted solitons still lie at peaks of the extended

background wave. (j) A disordered crystal. (k-n) Crystals exhibiting superstructure. (o) A crystal with irregular inter-soliton spacings. Darker shading indicates a smaller inter-soliton spacing. The range in inter-soliton spacings is 3 % of the mean. (p-t) A series of crystals generated as the pump laser is moved progressively closer to the stabilizing mode crossing.

The Frenkel defect, occurring once per round-trip, contributes the single-FSR lobes to the spectrum. The result is three bursts of 8, 9, and 10 solitons respectively.

Fig. 3o shows a soliton crystal with inter-soliton separations that are slightly irregular and which we have not simulated as a steady-state solution of any perturbed LLE. We expect that the stability of the crystal and the distribution of solitons are determined by mode-interactions, but that in this case our simple approximation of a perturbation to the LLE by a reduced comb-resonator detuning on a single comb mode is not appropriate.

**Time-domain measurements of Kerr soliton crystallization.** It is recognized in ultrafast optics that it is not generally possible to infer the time-domain waveform of an optical signal from its power spectrum without additional information[39]. So far in this paper we have assumed that the time-domain waveform must be a superposition of solitons, as this is a fundamental nonlinear-wave solution for Kerr combs. We can confirm this interpretation of the spectral data by performing intensity cross-correlation measurements between the soliton crystals and a single reference pulse; below we show a representative measurement for crystal 3j. For details see the Methods.

In Fig. 4, we plot the simulated time-domain waveforms of the reference pulse and the crystal (panel a), the simulated and measured cross-correlation signals (panel b), and the temporal spacing between peaks of the cross-correlation signals (panel c). We observe agreement between the measured and simulated cross-correlation signals, confirming both our generic interpretation of the spectrum as a superposition of solitons and the specific inversion of this spectrum to determine the distribution of the solitons in this disordered crystal. Beyond confirming our interpretation of the data, the cross-correlation measurement represents the incorporation of a completely new tool into the repertoire of techniques available for investigating Kerr combs.

## Discussion

Our discovery of spontaneous soliton crystallization in monolithic Kerr resonators advances the field of microcombs by exploring microcomb soliton physics in a new tightly packed regime, and it demonstrates that adiabatic soliton generation is possible in these systems with the help of avoided mode-crossings. A major benefit of using soliton crystals for microcomb applications is the enhanced efficiency $P_{comb}/P_{pump}$ of these combs, because $P_{comb}$ scales linearly with the number of solitons co-circulating in the resonator, each of which converts a fixed amount of power from the pump to the comb while preserving the level of the CW background. Soliton crystallization in Kerr resonators suggests a new route for customizability of frequency-comb spectra and time-domain waveforms through engineering of mode interactions[4,33]. Additionally, our work and other theoretical work[34,36] suggests that it may be possible to use a mode crossing or a bichromatic pump to generate a seed crystal for creating complex, custom soliton trains, which could be populated through the use of a pulsed pump laser. Potential applications of storage and manipulation of solitons in nonlinear resonators have attracted considerable interest[13–16,26,40], and soliton crystals in monolithic Kerr resonators present a possible mechanism for chip-integrated optical data storage and manipulation which exploits the enormous degeneracy of soliton crystal configuration space and is stable to perturbations and local interactions. Finally, the mechanism we present here for soliton-soliton interactions has implications for periodic systems in any context in which nonlinearity and dispersion or diffraction contribute significantly to the dynamics.

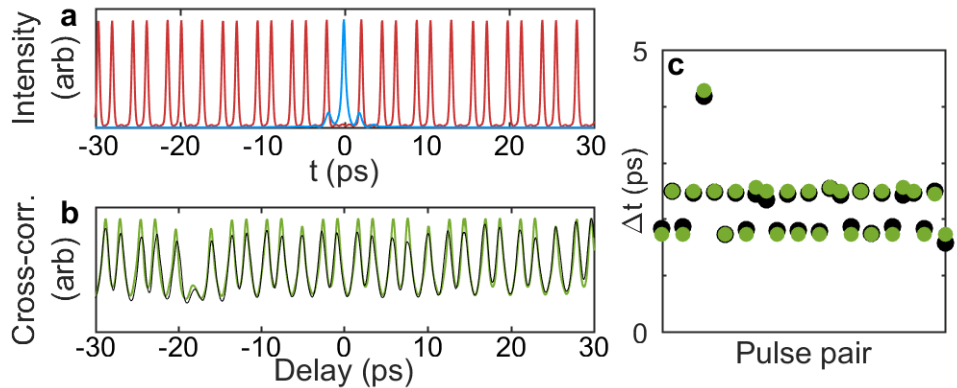

Figure 4. Intensity cross-correlation measurements of crystal (j) from Fig. 3. (a) Simulated crystal (red) and reference (blue) intensity profiles. (b) Measured (black) and simulated (green) cross-correlation signals. The contrast between peaks of the cross-correlation signal, for both theory and experiment, is limited by the duration and shape of the reference pulse, and increases between soliton pairs with larger temporal separations. (c) Temporal separations between adjacent peaks for the measured (black) and simulated (green) cross-correlation signals. Mean fractional error is 3.5 %.

# Methods

**Crystal generation:** Soliton crystals are generated in 16.5 GHz FSR disk[31] and 26 GHz FSR rod[30] silica resonators. The silica resonator has dispersion of ~5 kHz/mode; the rod resonator is described in Ref. 24. For crystal generation, a telecom-wavelength pump laser (typically 1550–1565 nm) is coupled into a whispering-gallery mode of the resonator using a tapered optical fiber[41] and scanned with decreasing optical frequency across the resonance. A sawtooth-shaped frequency scan is used, with a period between 10 ms and 1 s. The scan range is typically on the order of 3 GHz, for scan rates from 3 to 300 GHz/s. Soliton crystals may also be generated by scanning the pump frequency arbitrarily slowly by hand. The pump laser's power is set to between 2.5 and 6 times the measured absolute threshold power for parametric oscillation, which typically results in 30–80 mW power input to the taper. To measure the optical spectrum of primary comb, chaos, or soliton crystal, the frequency scan is stopped at an appropriate point where the desired state exists. During the laser scan, the resonator absorbs laser power and heats up, which leads to an increase in the resonant wavelength due to thermo-optic effects that increase the optical path length. This results in a triangular, rather than Lorentzian, lineshape as the laser frequency is decreased[32], a greatly increased resonant linewidth, and a hysteretic lineshape that is dependent upon the direction of the scan.

Diagnostics collected during crystal generation include the optical power transmitted through the tapered fiber, this same power with the pump frequency removed using a spatial-light modulator or optical band-reject filter, and the repetition-rate signal of the generated optical waveform collected by a 50 GHz photodiode.

Because crystal generation relies on a coupling between nominally orthogonal mode families in the resonator, it is sensitive to the position of the tapered optical fiber. The taper contacts the resonator and provides the coupling between mode families. The optimum taper position for crystal generation, in terms of position within the plane perpendicular to propagation and position along the taper's axis, is determined experimentally.

**Dynamical model:** Kerr comb generation is described by the Lugiato-Lefever equation[8,9,42]:

$$\frac{\partial \psi}{\partial \tau} = -(1+i\alpha)\psi + i|\psi|^2\psi - i\frac{\beta}{2}\frac{\partial^2 \psi}{\partial \theta^2} + F. \tag{1}$$

Here $\psi$ is the intracavity field, whose square-modulus is normalized to the threshold intracavity power[8] for parametric oscillation. The normalized time $\tau$ is equal to $t/2\tau_\gamma$, where $\tau_\gamma = 1/\Delta\omega$ is the resonator photon life-time. The pump detuning $\alpha$ and the dispersion $\beta$ are both normalized to half the resonance linewidth $\Delta\omega$: $\alpha = -2(\omega_p - \omega_0)/\Delta\omega$, where $\omega_0$ is the angular frequency of the pumped resonance and $\omega_p$ is the angular frequency of the pump laser; $\beta = -2\xi_2/\Delta\omega$, where $\xi_2 = \left.\frac{\partial^2 \omega_\mu}{\partial \mu^2}\right|_{\mu=0}$, $\omega_\mu$ being the angular frequency of the $\mu^{\text{th}}$ resonator mode relative to the pumped mode, for which $\mu = 0$. The intracavity angular coordinate $\theta$, which ranges from $0$ to $2\pi$, is measured in a frame which co-rotates at the group velocity of the pulses. The pump power $F^2$ is normalized to the absolute threshold for parametric oscillation: $F^2 = P_{\text{pump}}/P_{\text{threshold}}$.

**Numerical simulations:** Numerical simulations are dynamic simulations of the Lugiato-Lefever equation via an adaptive[43] Runge-Kutta in the interaction picture[44] (RK4IP) method. This is a Fourier split-step method, in which the dispersion operator is applied in the frequency domain. Periodic boundary conditions are implicitly taken into account through the use of the fast Fourier transform algorithm and the specification of the intracavity field for one resonator round trip. Our simulation method can equivalently be viewed as an application of the coupled-mode formalism[45], with the nonlinear term evaluated in the time domain for computational efficiency.

To simulate steady-state soliton crystals, initial conditions of the simulation must seed solitons in the appropriate locations, because solitons do not form spontaneously. Simulations of primary comb and chaos are run from zero initial intracavity field, and require the inclusion of simulated vacuum fluctuations. The simulated chaotic spectrum presented in Fig. 1c is a time-average, which is what is collected in the experiment due to the acquisition time of the optical spectrum analyzer—the chaotic spectrum varies on the time-scale of the photon life-time, as does the intracavity field, of which we have presented a simulated snapshot in Fig. 1d.

To perturb the LLE to account for the effect of a frequency shift of resonator modes due to a mode crossing, we note that it is possible to calculate the mode-dependent comb-resonator detuning $\alpha_\mu = -2(\omega_p + \xi_1\mu - \omega_\mu)/\Delta\omega$,

$\xi_1$ being the FSR of the resonator at the pumped mode $\xi_1 = \frac{\partial \omega_\mu}{\partial \mu}\Big|_{\mu=0}$, by grouping together the pump-detuning term with the Fourier-transformed dispersion term: $\alpha_\mu = \alpha - \beta\mu^2/2$, where $\beta < 0$ indicates anomalous dispersion and permits the formation of bright solitons. This formulation can be extended as necessary to include higher order dispersion, but we do not need to include higher-order dispersion in our simulations. Importantly, the formulation also permits inclusion of local changes in resonator mode structure through $\delta$-function perturbations of the comb-resonator detuning, in the form $\alpha_\mu = \alpha - \frac{\beta\mu^2}{2} + \Delta\alpha_\mu$, $\Delta\alpha_\mu$ being the normalized change in the frequency of the resonator mode: $\Delta\alpha_\mu = 2(\omega_\mu - \omega_{\mu,o})/\Delta\omega$.

In our implementation here, the LLE describes the evolution of the intracavity field. The experimental data we collect reflects the power spectrum of the field propagating away from the resonator in the tapered fiber, which includes through-coupled pump light. Thus, the relative amplitude of the pump laser with respect to the crystal is different for our experiments and our LLE simulations. We have corrected this by phenomenologically adjusting the pump laser power in the simulated optical spectrum to match the experimental data after the simulation is complete. In this way, we both account for the physical effect of the through-coupled pump and correctly simulate the soliton crystal dynamics, which the through-coupled pump does not affect.

**Simulation of soliton crystals:** The simulation of a soliton crystal exploits two independent experimental observations which are mutually confirmatory. First, the time-domain pulse train is deduced from the complicated shape of the optical spectrum of a soliton crystal by beginning with the assumption that the spectrum arises from a superposition of solitons, with no invocation of the crystallization theory we have presented above. A simple example is the pulse train corresponding to the "primary-comb-plus-soliton" spectrum in Fig. 1d, in which a soliton is added out of phase to a uniform soliton pulse train having a primary-comb-like spectrum. This soliton contributes the underlying single-FSR spectral envelope and eliminates a soliton from the pulse train in the time domain. Crystals with Frenkel defects may be understood in this way as well: the lobed structure in the spectrum is due to the spectral interference between an out-of-phase soliton which is added to the pulse train to remove a soliton, as above, and an in-phase soliton which is added to yield the shifted pulse.

Once the pulse train corresponding to the general structure of the spectrum has been deduced, the experimental spectrum can be compared to the spectrum of this pulse train calculated as a simple superposition of sech$^2$ pulses. This comparison reveals localized excess power in the experimental spectrum, which is evidence of a mode crossing. The strength of the mode crossing, represented in the simulation by $\Delta\alpha_{\mu_x}$ and controlled in the experiment by the taper-induced coupling between the modes, is determined phenomenologically from the magnitude of the excess power, and reduced comb-resonator detuning on a single optical mode is incorporated into a perturbed LLE. It is then verified that the pulse train whose spectrum matches the experimental data is a steady-state solution to this perturbed LLE, which requires the inter-soliton separations to be multiples of the period of the beat between the excess power and the pump laser. This period is $2\pi/\mu_x$, where $\mu_x$ is the mode number at the crossing. This process connects the presence of excess power on a single optical mode of the spectrum to the general shape of the spectrum, two observations which a priori are not related.

This process is applied for all of the crystal states presented in Fig. 3 except for the disordered crystals and the crystal 3o. For 10 of the 13 experimental crystal spectra to which we apply the model, the stabilizing mode crossing is visible in the data but not necessarily shown in the figure; for the other three the position of the mode crossing is inferred from the distribution of solitons and other crystal states observed in the same resonator. To model the disordered crystals, the pulse train is not deduced from the shape of the optical spectrum. Instead, excess power due to a mode crossing is identified through the presence of an asymmetry in the spectrum about the pump. The location of the excess power then fixes the allowed inter-soliton separations in the resonator, after which an exhaustive search is performed until a pulse train is found that yields the experimental spectrum. As described in the text, we have not simulated spectrum 3o as a steady-state solution to any perturbed LLE, which we expect is because the crystal is stabilized by a more complex spectrum of excess power.

**Instability of soliton crystals under the LLE:** None of the crystals we present are steady-state solutions to the unperturbed LLE. To arrive at this conclusion, we note that the temporal width of the pulses is determined by the bandwidth of the spectrum. This sets a lower bound on the range of their attractive interactions, which is too large for any of the crystals to be stable.

In Extended Data Fig. E1, we demonstrate the crystallization of an initially non-uniform pulse train due to the presence of a mode crossing, as well as the instability of the same pulse train without the mode crossing to stabilize it.

**Cross-correlation measurements:** Extended Data Fig. E2 presents a schematic depiction of the cross-correlation measurements, for which we use a commercial optical cross-correlator with a $LiNbO_3$ crystal. We use for a reference pulse the output of an electro-optic comb generator[46] whose repetition rate is locked to the repetition rate of the ~16.5 GHz crystal pulse train and which is generated from the same pump laser. When the reference pulse and the soliton crystal pulse train are sent together into the nonlinear crystal exhibiting the $\chi_2$ nonlinearity at 90º angles to each other, an amount of light proportional to the product of their intensities, at the sum of their frequencies, is emitted in a third direction. By measuring the average power of this emitted light while scanning the relative delay, we measure the intensity cross-correlation between the crystal and the reference pulse.

We operate our experiment in a through-coupled configuration, which results in destructive interference between the out-coupled solitons and the through-coupled pump. The solitons manifests as dips in the through-coupled intensity, resembling so-called dark solitons[8]. To correct this we use a spatial light modulator to rotate the phase of the pump laser by $\pi$ so that it constructively interferes with the solitons, yielding solitons riding atop a CW background. This CW background is larger than the CW background which exists inside the cavity because it arises from the pump laser which is transmitted past the resonator.

The simulated cross-correlation signal is sensitive to the intensity profile of the reference pulse. We can measure only its intensity autocorrelation, which we combine with our knowledge of its production to estimate the intensity profile. To demonstrate that the validity of the results we present here is not sensitive to the exact assumptions we make about the intensity profile, we have also simulated the intensity cross-correlation resulting from an assumed Gaussian reference pulse with the same autocorrelation width as is measured for the reference pulse. The resulting simulated cross-correlation does not qualitatively agree as well with the experimental data in the depths of the wells between peaks because it does not contain satellite pulses which contribute to the variations in this depth, but the quantitative comparison of the temporal spacing between peaks is similar: the mean (maximum) normalized error between experiment and theory is 3.5 % (9.1 %) for the assumed electro-optic comb pulse and 4.8 % (10.6 %) for the Gaussian pulse.

# References

1. Kippenberg, T. J., Holzwarth, R. & Diddams, S. A. Microresonator-based optical frequency combs. *Science* **332,** 555–559 (2011).

2. Diddams, S. A. The evolving optical frequency comb [Invited]. *J. Opt. Soc. Am. B* **27,** B51–B62 (2010).

3. Herr, T. *et al.* Temporal solitons in optical microresonators. *Nat. Photonics* **8,** 145–152 (2014).

4. Herr, T. *et al.* Mode Spectrum and Temporal Soliton Formation in Optical Microresonators. *Phys. Rev. Lett.* **113,** 123901 (2014).

5. Joshi, C. *et al.* Thermally Controlled Comb Generation and Soliton Modelocking in Microresonators. *Opt. Lett.* **41,** 2565–2568 (2016).

6. Yi, X., Yang, Q.-F., Yang, K. Y., Suh, M.-G. & Vahala, K. Soliton frequency comb at microwave rates in a high-Q silica microresonator. *Optica* **2,** 1078–1085 (2015).

7. Drake, T. E. *et al.* An octave-bandwidth Kerr optical frequency comb on a silicon chip. *Conf. Lasers Electro-Optics, Tech. Dig.* STu3Q.4 (2016).

8. Godey, C., Balakireva, I. V., Coillet, A. & Chembo, Y. K. Stability analysis of the spatiotemporal Lugiato-Lefever model for Kerr optical frequency combs in the anomalous and normal dispersion regimes. *Phys. Rev. A* **89,** 63814 (2014).

9. Coen, S., Randle, H. G., Sylvestre, T. & Erkintalo, M. Modeling of octave-spanning Kerr frequency combs using a generalized mean-field Lugiato-Lefever model. *Opt. Lett.* **38,** 37–39 (2013).

10. Coen, S. & Erkintalo, M. Universal scaling laws of Kerr frequency combs. *Opt. Lett.* **38,** 1790–2 (2013).

11. Yang, K. Y. *et al.* Broadband dispersion-engineered microresonator on a chip. *Nat. Photonics* **10,** 316–320 (2016).

12. Okawachi, Y. *et al.* Bandwidth shaping of microresonator-based frequency combs via dispersion engineering. *Opt. Lett.* **39,** 3535–3538 (2014).

13. Jang, J. K., Erkintalo, M., Coen, S. & Murdoch, S. G. Temporal tweezing of light through the trapping and manipulation of temporal cavity solitons. *Nat Commun* **6,** 1–7 (2015).

14. Leo, F. *et al.* Temporal cavity solitons in one-dimensional Kerr media as bits in an all-optical buffer. *Nat. Photonics* **4,** 471–476 (2010).

15. Jang, J. K. *et al.* All-optical buffer based on temporal cavity solitons operating at 10 Gb / s. **41,** 4526–4529 (2016).

16. Jang, J. K., Erkintalo, M., Murdoch, S. G. & Coen, S. Ultraweak long-range interactions of solitons observed over astronomical distances. *Nat. Photonics* **7,** 657–663 (2013).

17. Malomed, B. A., Schwache, A. & Mitschke, F. Soliton lattice and gas in passive fiber-ring resonators. *Fiber Integr. Opt.* **17,** 267–277 (1998).

18. Mitschke, F. & Schwache, A. Soliton ensembles in a nonlinear resonator. *J. Opt. B Quantum Semiclassical Opt.* **10,** 779–788 (1998).

19. Schwache, a. & Mitschke, F. Properties of an optical soliton gas. *Phys. Rev. E* **55,** 7720–7725 (1997).

20. Zajnulina, M. *et al.* Characteristics and stability of soliton crystals in optical fibres for the purpose of optical frequency comb generation. *Opt. Commun.* **393,** 95–102 (2017).

21. Haboucha, A., Leblond, H., Salhi, M., Komarov, A. & Sanchez, F. Coherent soliton pattern formation in a fiber laser. *Opt. Lett.* **33,** 524 (2008).

22. Amrani, F., Salhi, M., Grelu, P., Leblond, H. & Sanchez, F. Universal soliton pattern formations in passively mode-locked fiber lasers. *Opt. Lett.* **36,** 1545–7 (2011).

23. Haboucha, A., Leblond, H., Salhi, M., Komarov, A. & Sanchez, F. Analysis of soliton pattern formation in passively mode-locked fiber lasers. *Phys. Rev. A* **78,** 1–12 (2008).

24. Ashcroft, N. W. & Mermin, D. N. *Solid State Physics*. (Brooks/Cole, 1976).

25. Herr, T. *et al.* Universal formation dynamics and noise of Kerr-frequency combs in microresonators. *Nat. Photonics* **6,** 480–487 (2012).

26. McDonald, G. S. & Firth, W. Spatial solitary wave optical memory. *J. Opt. Soc. Am. B* **7,** 1328–1335 (1990).

27. Luo, K., Jang, J. K., Coen, S., Murdoch, S. G. & Erkintalo, M. Spontaneous creation and annihilation of temporal cavity solitons in a coherently driven passive fiber resonator. *Opt. Lett.* **40,** 3735–3738 (2015).

28. Del'Haye, P., Beha, K., Papp, S. B. & Diddams, S. A. Self-injection locking and phase-locked states in microresonator-based optical frequency combs. *Phys. Rev. Lett.* **112,** 43905 (2014).

29. Del'Haye, P. *et al.* Phase steps and resonator detuning measurements in microresonator frequency combs. *Nat. Commun.* **6,** 5668 (2015).

30. Del'Haye, P., Diddams, S. A. & Papp, S. B. Laser-machined ultra-high-Q microrod resonators for nonlinear optics. *Appl. Phys. Lett.* **102,** 221119 (2013).

31. Lee, H. *et al.* Chemically etched ultrahigh-Q wedge-resonator on a silicon chip. *Nat. Photonics* **6,** 369–373 (2012).

32. Carmon, T., Yang, L. & Vahala, K. J. Dynamical thermal behavior and thermal self-stability of microcavities. *Opt. Express* **12,** 4742–4750 (2004).

33. Liu, Y. *et al.* Investigation of mode coupling in normal-dispersion silicon nitride microresonators for Kerr frequency comb generation. *Optica* **1,** 137–144 (2014).

34. Hansson, T. & Wabnitz, S. Bichromatically pumped microresonator frequency combs. *Phys. Rev. A* **90,** 13811 (2014).

35. Skryabin, D. V & Firth, W. J. Interaction of cavity solitons in degenerate optical parametric oscillators. *Opt. Lett.* **24,** 1056–8 (1999).

36. Wabnitz, S. Control of soliton train transmission, storage, and clock recovery by cw light injection. *J. Opt. Soc. Am. B Opt. Phys.* **13,** 2739–2749 (1996).

37. Haus, H. A. & Huang, W. Coupled-Mode Theory. *Proc. IEEE* **79,** 1505–1518 (1991).

38. Ibach, H. & Luth, H. *Solid-State Physics: An Introduction to Principles of Materials Science*. (Springer,

2009).

39. Weiner, A. *Ultrafast Optics*. (Wiley, 2009).

40. Pang, M., He, W., Jiang, X. & Russell, P. S. J. All-optical bit storage in a fibre laser by optomechanically bound states of solitons. *Nat. Photonics* **10,** 454–458 (2016).

41. Spillane, S. M., Kippenberg, T. J., Painter, O. J. & Vahala, K. J. Ideality in a fiber-taper-coupled microresonator system for application to cavity quantum electrodynamics. *Phys. Rev. Lett.* **91,** 43902 (2003).

42. Chembo, Y. K. & Menyuk, C. R. Spatiotemporal Lugiato-Lefever formalism for Kerr-comb generation in whispering-gallery-mode resonators. *Phys. Rev. A* **87,** 53852 (2013).

43. Sinkin, O. V., Holzlöhner, R., Zweck, J. & Menyuk, C. R. Optimization of the split-step Fourier method in modeling optical-fiber communications systems. *J. Light. Technol.* **21,** 61–68 (2003).

44. Hult, J. A Fourth-Order Runge-Kutta in the Interaction Picture Method for Simulating Supercontinuum Generation in Optical Fibers. *J. Light. Technol.* **25,** 3770–3775 (2007).

45. Chembo, Y. ~K. K. & Yu, N. Modal expansion approach to optical-frequency-comb generation with monolithic whispering-gallery-mode resonators. *\Pra* **82,** 33801 (2010).

46. Beha, K. *et al.* Self-referencing a continuous-wave laser with electro-optic modulation. *arXiv: 1507.06344* (2015).

## Supplementary Information

**Phase steps:** Soliton crystals exhibit phase steps previously reported by Del'Haye et al.[29] We plot simulations of several crystals with spectral phase in Extended Data Fig. E3. We note that, because a linear spectral phase shift is equivalent to a shift in time, there is an infinite continuum of ways to present the spectral phase for a given crystal, and they may not appear equivalent. The phase steps exhibited by soliton crystals arise due to the superposed linear spectral phase shifts from temporally-separated co-propagating solitons.

**Harmonic mode-locking:** We have compared and contrasted soliton ensembles in the tightly-packed and sparsely-populated regimes in the main text. In conducting numerical simulations, we have also observed harmonic mode-locking in Kerr combs. Harmonic mode-locking consists of the generation of a uniform soliton train, and occurs in a density regime between the low-density and high-density regimes discussed in the text. In simulations, soliton ensembles in an intermediate-density regime which are initialized with non-uniform soliton distributions will evolve to uniformity, with a spectrum resembling primary comb. Harmonic mode-locking does not require the presence of a mode crossing, and the harmonic mode-locking we have observed involves only the evolution towards a uniform pulse train; no other crystal structure has been discovered. Harmonic mode-locking is a weak effect: the timescale over which a non-uniform pulse distribution will evolve to uniformity in the case of harmonic mode-locking is on the scale of $10^3$–$10^5$ photon lifetimes, compared with <10 photon lifetimes for crystallization of a non-uniform pulse train under the influence of a mode crossing. In Extended Data Fig. E4 we present simulations of harmonic mode-locking beginning from a uniform pulse train with jittered pulse positions, and beginning from a uniform pulse train with a vacancy. The numbers of solitons in these harmonically mode-locked states are 15 and 14, respectively; the simulation is conducted under the same conditions in which we have experimentally observed crystallization with 23 pulses in the presence of a mode crossing.

# Extended Data

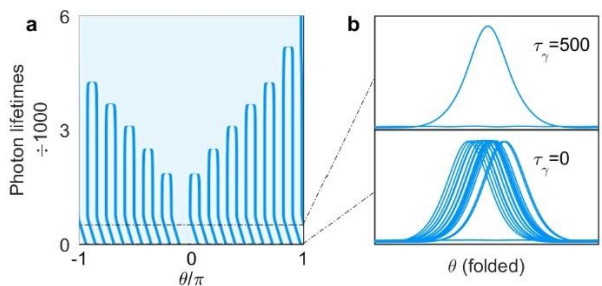

Fig. E1. (a) Simulated evolution of the pulse train corresponding to the experimental crystal spectrum shown in Fig. 1, starting from irregular pulse positions. Crystallization occurs within 10 photon lifetimes of the initialization of the simulation. For the first 500 photon lifetimes of the simulation, the propagation is governed by a perturbed LLE including reduced comb-resonator detuning on modes 5 x 24=120 and 7 x 24=168. During this time the crystal drifts within the co-rotating frame because the

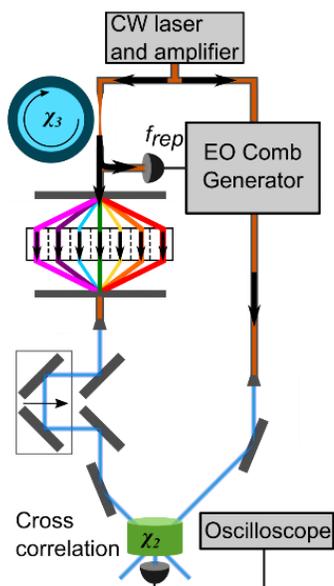

optical spectrum is asymmetric. The perturbation is then removed smoothly from 500 to 1000 photon lifetimes and subsequently the pulses pair-annihilate, demonstrating that the mode crossing is critical for stabilizing the soliton crystal. (b) Intracavity power with the angular coordinate folded modulo $2\pi/24$, to depict the irregularity of the pulse positions at the initialization of the simulation and the crystallized pulse train after 500 photon lifetimes.

Fig. E2. Schematic depiction of the set-up for using an electro-optic (EO) modulator comb as a reference pulse to measure the time-domain waveform of the crystal. The $\chi_3$ (Kerr) and $\chi_2$ nonlinearities are indicated on the resonator and nonlinear crystal, respectively. A spatial-light modulator is used to rotate the phase of the pump laser by $\pi$ in the crystal to improve the cross-correlation contrast. The soliton crystal and the EO modulator comb share a pump laser, and the repetition frequency $f_{\text{rep}}$ of the EO modulator comb is locked to that of the crystal. Varying the relative delay in one arm of the interferometer allows a measurement of the intensity cross-correlation between the crystal and the reference pulse.

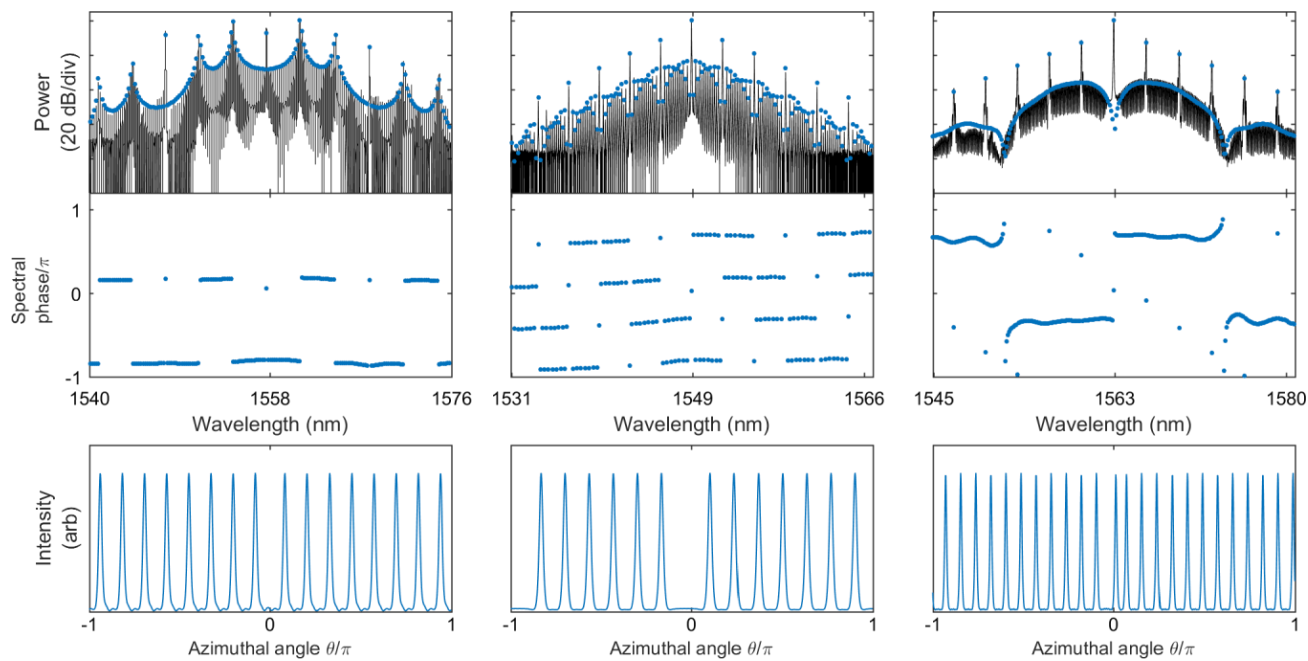

Fig E3. Depiction of crystals (p), (d), and (g) from Fig. 3 of the main text, with experimental data in black and simulations in blue, including simulated spectral phase. These crystals exhibit phase steps near prominent features of the spectrum, which arise due to the spectral interference between temporally-shifted solitons. Spectral phase is unique only up to a shift linear in frequency, which can yield multiple phase profiles for the same crystal which appear very different.

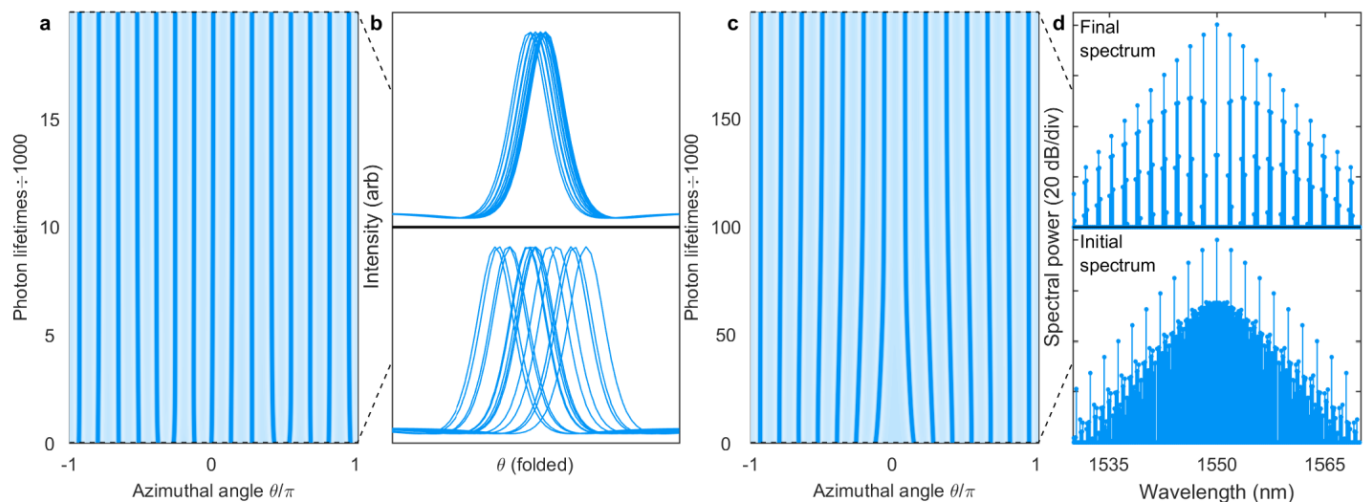

Fig E4. Simulations of harmonic mode-locking, or the evolution towards a uniform soliton train without the presence of a mode crossing. (a-b) Harmonic mode-locking of 15 pulses beginning from jittered, non-uniform pulse positions. Initial and final pulse configurations are shown in (b) with the angular coordinate folded modulo $2\pi/15$ to illustrate the degree of regularity of the pulse distribution. The pulse distribution evolves towards a uniform pulse train, but approaches that distribution slowly. (c-d) Harmonic mode-locking of 14 pulses beginning from a uniform 15-soliton pulse train with a single vacancy. Initial and final spectra shown in (d).

## Acknowledgements

We thank Daniel Hickstein and Kyle Beloy for comments on the manuscript, and Ki Youl Yang and Kerry Vahala for providing the 16 GHz wedge resonators. This material is based upon work supported by the Air Force Office of Scientific Research under award number FA9550-16-1-0016. Additional support is provided by the NIST-on-a-Chip program and the DARPA QuASAR and PULSE programs. DC acknowledges support from the NSF GRFP under

Grant No. DGE 1144083. This work is a contribution of the US government and is not subject to copyright in the United States of America.